# Trends and Techniques in Visual Gaze Analysis

Sophie Stellmach[1], Lennart Nacke[2], Raimund Dachselt[1], and Craig A. Lindley[2]

[1]User Interface & Software Engineering Group, Department of Simulation and Graphics, Faculty of Computer Science, Otto-von-Guericke University
PO Box 4120, D-39016 Magdeburg, Germany
sstellmach@gmail.com, dachselt@acm.org

[2]Game and Media Arts Laboratory, Department of Interaction and System Design, School of Engineering, Blekinge Institute of Technology
PO Box 214, SE-374 24 Karlshamn, Sweden
Lennart.Nacke@bth.se, Craig.Lindley@bth.se

**Keywords**
Gaze visualization, trends, virtual environments, qualitative and quantitative survey

## Introduction

Visualizing gaze data is an effective way for the quick interpretation of eye tracking results. In the two general application areas of eye tracking, diagnostics and interaction (Duchowski, 2002), recently there has been put much effort into gaze interaction for three-dimensional (*3D*) virtual environments (*VEs*) (Castellina and Corno, 2008; Isokoski and Martin, 2006; Smith and Graham, 2006). However, since diagnostic studies benefit from visualizations of eye tracking data for understanding complex relationships between gaze behavior and stimuli, developing visualization techniques for *3D VEs* is a fundamental next step in eye tracking research.

A classification of gaze visualization techniques by Špakov (Špakov, 2008, pp. 37-49) emphasized the limited variety of suitable techniques for *3D* stimuli. The most widely used procedure for investigating gaze data for dynamic and *3D* stimuli is to analyze superimposed gaze plots over video recordings on a frame-by-frame basis. This quickly results in a monotonous and time-consuming process. The lack of suitable techniques for a more efficient gaze analysis of *3D VEs* results in the desire for enhanced procedures. Such techniques may provide quick insights into gaze behavior for evaluative studies of, for example, digital games, model designs, and product placement in virtual worlds. The purpose of the research presented here is to establish a foundation for improving gaze visualizations of eye tracking data. We conducted a survey with professionals and researchers working with different stimulus types to find out more about the importance of gaze visualizations and general requirements for improved eye tracking analysis. This research aims at gaining a formal understanding of gaze visualization techniques and applying this knowledge to the design and development of novel visualizations especially for *VEs*.

In this paper, preliminary findings from mixed-method (some quantitative, but a major emphasis on qualitative) questions will be presented and discussed in light of the proposed gaze visualization framework. It has to be noted that this was primarily a qualitative investigation, which explains the small and *not* randomly selected sample size, thus following the purposeful sampling strategy discussed by Creswell (2007, pp. 125-129).

## Method

**Participants.** Ten eye tracking professionals and researchers aged between 28 and 57 years (*Mean (M)* = 37.7, *Standard Deviation (SD)* = 9.38) participated in this survey online. Of the total, 50% (*N* = 5) were female and 50% (*N* = 5) were male. Participants had been working with eye trackers between 2 and 15 years with an average of 7.2 years (*SD* = 4.49). Having to grade their knowledge concerning eye tracking on a scale from 1 (little knowledge) to 5 (much knowledge), the average knowledge of participants was high ($M = 4.2$, $SD = 0.63$). When asked how many studies they had worked on that incorporated eye tracking analysis, participants' experience ranged between 3 and 50 studies ($M = 15.9$, $SD = 14.71$).

**Survey design, apparatus and procedure.** The survey consisted of demographic, mixed-method (eight quantitative and four qualitative) questions. The quantitative questions were aimed at evaluating the importance of visualizations for eye tracking analysis ("*How important were visualizations for the analysis of your eye tracking studies?*", "*How would you assess the importance of sophisticated gaze visualization techniques for dynamic virtual environments?*"). These evaluations were done on a scale from 1, not important, to 5, very important. Other quantitative questions were aimed at uncovering the stimuli types used in these studies (static [two-dimensional (*2D*), *3D*], dynamic [*2D*, *3D*], interactive [*2D*, *3D*]). The qualitative questions asked about personal experiences ("*What are your experiences in designing and analyzing eye tracking experiments employing dynamic interactive stimuli?*"), weaknesses of current gaze visualizations ("*Where do you see weaknesses in current gaze visualization techniques?*") and desirable features for gaze analysis ("*What features would you desire for a simple and intuitive gaze analysis?*"). The survey was implemented online using the tool LimeSurvey[1] (Version 1.70+). Thirty-two researchers were selected from searches on major eye tracking publication venues (*COGAIN*, *ETRA* and *ECEM*) and together with staff from *Tobii Technology AB*. Anonymous identifiers were assigned to each participant and they were then invited via email to participate in the online survey. No financial incentives were offered for participation.

## Preliminary results

**Quantitative results.** Visualizations for eye tracking analysis as conducted by individual researchers were assessed as important ($M = 3.7$, $SD = 0.95$). Gaze visualizations for dynamic *VEs* were estimated to be a little more important ($M = 3.9$, $SD = 0.88$), although not significant, $t(9)=-0.56$, $p>.05$. Ninety percent of the participants had already used static *2D* stimuli in their experiments. Following this, 70% had drawn on interactive *2D* stimuli (user interfaces) and 40% had employed *2D* dynamic stimuli (videos). However, only 20% had already used *3D* stimuli of all kinds (static, dynamic and interactive), see also Figure 1.

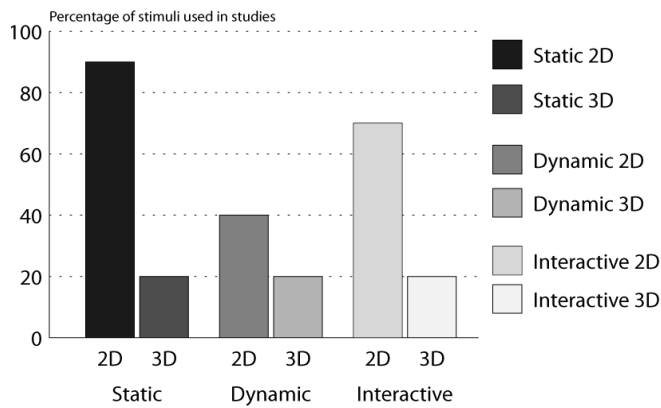

**Figure 1.** Percentages of stimuli (sorted by types) used by researchers in the study.

**Qualitative results.** Among the ten people interviewed, six people made a statement about their application areas (multiple selections were possible): four employed eye tracking for diagnosis and four for interactive applications. Two respondents had conducted studies using virtual *3D* stimuli.

---

[1] Open source survey application mainly developed by Carsten Schmitz and available for download at www.limesurvey.org

Although gaze visualizations are generally considered important ($M = 3.7$, $SD = 0.95$), researchers using eye tracking for interaction studies declared that they "*almost never did any visualizations*" (10 years work experience with eye tracking). For diagnostic purposes, however, visualizations provide "*the only way to quickly review and analyze what participants saw / did not see*" (8 years[2]). Thus, gaze visualizations can play an essential role in detecting gaze patterns. This is especially beneficial when trying to communicate findings to customers. Nevertheless, one respondent argued that visualizations "*do not prove anything*" (8 years).

In spite of the importance granted to gaze visualizations by most interviewees, several drawbacks were also specified. An important issue is the lack of effective dynamic visualization methods. In fact, eight people agreed with the following statement: "*So far, nobody has come across a simple and intuitive way to visualize gaze data for dynamic stimuli.*" (Ramloll et al., 2004; Špakov, 2008). One person disagreed, claiming that gaze replays are an intuitive way to visualize gaze positions. Another participant named several approaches for improving gaze visualizations for videos (Duchowski and McCormick, 1998; Nikolov et al., 2004; Sadasivan et al., 2005), but granted the difficulty of considering these techniques simple or intuitive.

A general problem for gaze visualizations is the danger of misinterpretation, which was mentioned by one interviewee. Reports on weaknesses of current visual gaze analyses mainly targeted two areas: features of available tools and of visualizations. Gaze analysis tools were criticized for the lack of multiple views and properly formatted data. In addition, further improvements of fixation identification algorithms and the conversion of processed visualizations into table data was demanded. It was also mentioned that gaze visualizations for *3D* contexts need to be improved and should provide aggregated comparisons (i.e. *3D* heat maps). Some participants warned that scan paths from multiple participants (especially in *3D*) may result in visual clutter as well as problems with temporal synchronization. Another respondent advised that "*simple gaze path animations may not help in seeing repeating patterns*" (8 years).

Concerning eye tracking studies within *VEs*, one interviewee requested the possibility to "*compare multiple scan paths in the VE while still allowing the person viewing this to control their own position/orientation in the VE*" (15 years). Another respondent (10 years) provided the following recommendations: "(*1*) *Easy way to specify the 'variables-of-interest' (not of AOIs[3]).(2) Multiple visualization and easy-to-follow links between them. (3) Ability to select and edit gaze data in a very flexible way (in tables, interactive, etc), having the variables-of-interest recalculating and multiple visualization repainting 'on-the-fly'*". Further requests include:

- Integration of signal processing methods like frequency analysis for automatic pattern detection
- Automatic relocation of *AOIs* for moving objects (videos)
- Visualization of transitions between different views
- Integration of external data sources
- Multiple views, overview-and-detail, backtracking
- Data format that allows use of statistical programs
- Annotation tool (allowing researchers to attach and share comments in the playback)
- Not being time consuming to set up and analyze (quick deployment)

## Discussion & Future Work

This paper described current trends and needs within eye tracking research concerning improvements in gaze analysis techniques. The presented results clearly imply the importance of gaze visualizations for diagnostic use. Aggregated visualizations such as heat maps may be integrated for static *3D* scenes. A very interesting result is that a majority of eye tracking researchers have not used *3D* stimuli so far. Reasons for this may include a higher complexity to develop sophisticated *3D* scenarios and the use of tedious frame-by-frame evaluation of session videos. Beside the adaption of visualizations to stimuli, several features concerning usability of available analysis tools have been mentioned, including: overview, details-on-demand, backtracking, etc. (see also the *Information Seeking Mantra,* (Shneiderman, 1996)).

---

[2] In the following X years denotes X years of work experience with eye tracking.

[3] Areas of interest

In conclusion, by identifying the trends and requirements for gaze visualization in *VEs*, we now have specifications for developing tools that can appropriately visualize various stimulus types in *VEs*. Our future work includes further investigation of visual analysis techniques for eye tracking as well as the prototyping of a tool that incorporates knowledge gained from this survey. The prototype tool will include the design and development of novel gaze visualizations for static *3D VE*s.

## Acknowledgements

This research was supported by "*FUGA* - The Fun of Gaming: Measuring the Human Experience of Media Enjoyment", funded under the 6th Framework Programme of the European Commission (Contract: *FP6-NEST-28765*). We thankfully acknowledge the participation of the eye tracking professionals and researchers in this study. Also, we would like to thank Jessica Lundqvist from *Tobii Technology AB* for recommending some of her contacts for participation in this study.